\documentclass[preprint,prd,aps,superscriptaddress,11pt]{revtex4-1}

\usepackage{graphicx}
\usepackage{amsmath}
\usepackage{amssymb}
\usepackage{dcolumn}
\usepackage{bm}
\usepackage{slashed}

\newcommand{\be}{\begin{equation}}
\newcommand{\ee}{\end{equation}}
\newcommand{\ba}{\begin{eqnarray}}
\newcommand{\ea}{\end{eqnarray}}
\newcommand{\no}{\nonumber \\}
\newcommand{\gsim}{\mathrel{\hbox{\rlap{\lower.55ex \hbox {$\sim$}}
                   \kern-.3em \raise.4ex \hbox{$>$}}}}
\newcommand{\lsim}{\mathrel{\hbox{\rlap{\lower.55ex \hbox {$\sim$}}
                   \kern-.3em \raise.4ex \hbox{$<$}}}}

\def\roughly#1{\mathrel{\raise.3ex\hbox{$#1$\kern-.75em%
\lower1ex\hbox{$\sim$}}}}
\def\lsim{\roughly<}
\def\gsim{\roughly>}

\def\({\left(}
\def\){\right)}
\def\[{\left[}
\def\]{\right]}
\def\<{\langle}
\def\>{\rangle}

\def\l{{\lambda}}

\def\d{{\delta}}
\def\D{{\Delta}}

\def\e{{\epsilon}}

\def\a{{\alpha}}
\def\b{{\beta}}

\def\P{{\Pi}}
\def\m{{\mu}}
\def\n{{\nu}}
\def\r{{\rho}}
\def\s{{\sigma}}

\def\P{{\Pi}}
\def\hb{{\hbar}}

\newcommand{\pd}{{\partial}}

\newcommand{\nab}{\Delta}

\date{\today}

\begin{document}

\title{\bf Photon Self-energy in Magnetized Chiral Plasma from Kinetic Theory}

\author{Han Gao}
\email{gaoh26@mail2.sysu.edu.cn}
\affiliation{School of Physics and Astronomy, Sun Yat-Sen University, Zhuhai 519082, China}
\author{Zonglin Mo}
\email{mozlin@mail2.sysu.edu.cn}
\affiliation{School of Physics and Astronomy, Sun Yat-Sen University, Zhuhai 519082, China}
\author{Shu Lin}
\email{linshu8@mail.sysu.edu.cn}
\affiliation{School of Physics and Astronomy, Sun Yat-Sen University, Zhuhai 519082, China}

\begin{abstract}
We study the photon self-energy in magnetized chiral plasma by solving the response of electromagnetic field perturbations in chiral kinetic theory with Landau level states. With lowest Landau level approximation and in collisionless limit, we find solutions for three particular perturbations: parallel electric field, static perpendicular electric and magnetic field, corresponding to chiral magnetic wave, drift state and tilted state, from which we extract components of photon self-energy in different kinematics. We show no solution is possible for more general field perturbations. We argue this is an artifact of the collisionless limit: while static solution corresponding to drift state and tilted state can be found, they cannot be realized dynamically without interaction between Landau levels. We also discuss possible manifestation of side-jump effect due to both boost and rotation, with the latter due to the presence of background magnetic field.

\end{abstract}

\maketitle

%\vspace{0.1in}

\newpage

\section{Introduction}

There has been a long history of effort towards understanding of vacuum polarization by electromagnetic fields. The full effective action of vacuum for arbitrary constant electromagnetic field was established by Heisenberg and Euler \cite{Heisenberg:1935qt}, which predicted critical electric field in vacuum. It was later realized by Schwinger \cite{Schwinger:1951nm} that the critical electric field leads to pair production. On the other hand, while the magnetic field does not destabilize the vacuum, it does modify vacuum properties: such as enhancing the pair production rate \cite{Dunne:2004nc} and causing vacuum birefringence \cite{Hattori:2012je,Hattori:2012ny}.

Recently there has been growing interests in the effect of magnetic field in chiral medium in a variety of systems including quark-gluon plasma and Weyl semimetal etc. The magnetic field in chiral medium is known to lead to novel anomalous transport such as chiral magnetic effect \cite{Vilenkin:1980fu,Kharzeev:2004ey,Kharzeev:2007tn,Fukushima:2008xe}, chiral separation effect \cite{Metlitski:2005pr,Son:2004tq} and chiral magnetic wave \cite{Kharzeev:2010gd} etc. Furthermore, the presence of magnetic field also modifies existing transport coefficient like conductivities \cite{Son:2012bg,Hattori:2016cnt,Hattori:2016lqx,Fukushima:2017lvb,Fukushima:2019ugr,Li:2018ufq,Lin:2019fqo,Astrakhantsev:2019zkr} and viscosities \cite{Critelli:2014kra,Li:2017tgi} nontrivially. In the regime of linear response, these transport phenomena are characterized by photon self-energy in magnetized chiral medium. In the presence of magnetic field, the photon self-energy contains very rich structure and is also very complicated in general. There have been many field theoretic attempts to study the photon self-energy \cite{Danielsson:1995rh,Chao:2014wla,Fukushima:2015wck,Chao:2016ysx}, see also studies on gluon self-energy \cite{Hattori:2017xoo,Singh:2020fsj}.

A distinguishing feature of chiral fermion from classical particle is its spin, which is a genuine quantum quantity measured in $\hb$. A semi-classical expansion in $\hb$ gives rise to the chiral kinetic theory (CKT) \cite{Son:2012wh,Son:2012zy,Stephanov:2012ki,Gao:2012ix,Pu:2010as,Chen:2012ca,Hidaka:2016yjf,Manuel:2013zaa,Manuel:2014dza,Wu:2016dam,Mueller:2017arw,Mueller:2017lzw,Huang:2018wdl,Gao:2018wmr,Carignano:2018gqt,Lin:2019ytz,Carignano:2019zsh,Liu:2018xip,Dayi:2018xdy,Weickgenannt:2019dks,Gao:2019znl,Hattori:2019ahi,Wang:2019moi,Gao:2019zhk,Liu:2020flb,Yang:2020hri}. It has been successfully applied to study transport phenomena of chiral medium in response to weak electromagnetic field \cite{Gorbar:2016qfh,Chen:2016xtg,Hidaka:2017auj,Abbasi:2018zoc}, where each power of electromagnetic field contributes $O(\hb)$. In the regime of strong magnetic field, a different expansion scheme is used giving rise to a chiral kinetic theory based on Landau level (LL) states \cite{Hattori:2016lqx,Sheng:2018jwf,Lin:2019fqo}. The purpose of this paper is to apply this chiral kinetic theory to study photon self-energy in magnetized chiral plasma as an alternative approach to the self-energy problem. For simplicity, we work in the strong magnetic field and collisionless limit. We will reproduce the field theoretic results to the accuracy of CKT approach and show some new results for drift state.

The paper is organized as follows: in Section II, we summarize generalities of photon self-energy; in Section III, we give a short review of chiral kinetic equations based on LL states and analyze the structure of equations; in Section IV, we present solutions corresponding to three specific perturbations and discuss the physical implications of them. We further show it is not possible to obtain more solution for more general perturbations. We will argue it is an artifact of collisionless limit; in Section V, we summarize the results and discuss future directions.

Throughout this paper, we primarily study chiral medium consisting of right-handed fermions with charge $Q=|e|$. Contribution of left-handed fermions will be added when comparing with field theoretic results. For simplicity, we set $e=1$ and reinstate it in the end. We use the following non-standard convention $p_\m=(p_0,p_1,p_2,p_3)$ for convenience. The magnetic field points in $x_3$ direction.

\section{Generalities of Photon self-energy in magnetized medium}

The photon self-energy in imaginary time formalism is defined by \cite{Bellac:2011kqa}
\begin{align}\label{PiE_x}
  \P^{\m\n}_E(x,x')=\<\hat{T}\(J_E^\m(x) J_E^\n(x')\)\>=\frac{\d J_E^\m(x)}{\d A^E_\n(x')},
\end{align}
where $\hat{T}$ denotes time ordering in Euclidean time. We can rewrite \eqref{PiE_x} in a simpler form in momentum space:
\begin{align}\label{PiE_p}
  \d J^\m_E(q)=\P^{\m\n}_E(q)\d A^E_\n(q).
\end{align}
Note that the Euclidean frequency $q_4$ takes discrete values of Matsubara frequencies $2\pi nT$, with $T$ being temperature. It can be analytically continued to complex frequency plane. Taking $q_4\to i(q_0+i\e)$, we obtain the more useful retarded photon self-energy
%\begin{align}\label{Pi_x}
%  \P^{\m\n}_R(x,x')=i\<[J(x)^\m,J(x')^\n]\>\th(t-t')=i\frac{\d J(x)^\m}{\d A(x')_\n}.
%\end{align}
%We choose to work with retarded self-energy, which can be studied from response of current to perturbation of electromagnetic field, as indicated above. In momentum space, we have
\begin{align}\label{Pi_p}
  \d J^\m(q)=\P^{\m\n}_R(q)\d A_\n(q),
\end{align}
with $J^4_E(q)\to iJ^0(q)$ and $A^4_E(q)\to iA^0(q)$. \eqref{Pi_p} expresses current as a response to external electromagnetic field perturbation, which can be studied in kinetic theory. We will mainly use the retarded self-energy in the paper.

For parity breaking chiral medium consisting of right handed fermions, $\P_R^{\m\n}$ is in general not symmetric in the Lorentz indices. Nevertheless, $\P_R^{\m\n}$ is still constrained by anomalous Ward identity. To derive anomalous Ward identity in the regime of strong magnetic field, we note that there is effective dimensional reduction from $3+1D$ to $1+1D$. In this regime, the most interesting perturbations are time and longitudinal components of photon field $\d A_0\equiv a_0$ and $\d A_3\equiv a_3$. For right handed current in the background magnetic field, the Ward identity is given by
\begin{align}\label{WI}
  \pd_\m J^\m=\frac{1}{(2\pi)^2}E_3B.
\end{align}
Here $E_3=\pd_0a_3-\pd_3a_0$ is electric field induced by perturbations. Fourier transforming \eqref{WI} and doing variation with respect to $a_0$ and $a_3$, we obtain the following constraints:
\begin{align}\label{aWI}
  &q_\m\P_R^{\m0}=\frac{1}{(2\pi)^2}(-q_3)B, \no
  &q_\m\P_R^{\m3}=\frac{1}{(2\pi)^2}q_0B.
\end{align}
Note that the anomalous Ward identity \eqref{aWI} involves all components of self-energy. We will use chiral kinetic theory to study the response.

\section{Chiral kinetic equations with Landau levels}

The chiral kinetic equations with Landau level states are given by \cite{Lin:2019fqo}
\begin{align}\label{ckt_LL}
  &\nab_0j^0+\nab_ij^i=0,\no
  &p_0j^0+p_ij^i=0,\no
  &\nab_0j^i+\nab_ij^0+2\e^{ijk}p_jj^k=0,\no
  &-p_0j^i-p_ij^0+\frac{1}{2}\e^{ijk}\nab_jj^k=0,
%  &\D_\m j^\m=0,\no
%  &-\D_\m j_\n+\D_\n j_\m+2\e_{\m\n\r\s}p^\r j^\s=0,\no
%  &p_\m j^\m=0, \no
%  &p_\m j_\n-p_\n j_\m+\frac{1}{2}\e_{\m\n\r\s}\D^\r j^\s=0,
\end{align}
with $\nab_\m=\pd^X_\m-(F_{\m\n}+f_{\m\n})\frac{\pd}{\pd p_\n}$ for $\m=0,1,2,3$. We use Greek letters for spacetime indices and small Roman letters for spatial indices. $F_{\m\n}$ corresponds to background magnetic field with the only nonvanishing components $F_{12}=-F_{21}=-B$. \eqref{ckt_LL} is derived based on an expansion in $\hb$, or equivalently in $\pd_X$. It is valid up to $O(\pd_X)$ and to all order in $B$, which implicitly assumes the hierarchy of scales $\pd_X\ll p\sim \sqrt{B}$. $f_{\m\n}=\pd^X_\m a_\n-\pd^X_\n a_\m$ corresponds to perturbation of electromagnetic field, which is counted as $O(\pd_X)$. Solving \eqref{ckt_LL} we can obtain $j^\m$ and the momentum integral of $j^\m$ gives the induced current $J^\m$%\footnote{The minus sign is due to our non-standard convention}:
\begin{align}\label{J}
  J^\m(X)=\int d^4p j^\m(X,p).
\end{align}
In the absence of perturbation, the background in the LLL approximation is given by
\begin{align}\label{bkg}
  j^0=j^3=\frac{2}{(2\pi)^3}\d(p_0+p_3)e^{-p_T^2/B}f(|p_0|),\quad j^1=j^2=0.
\end{align}
%Note minus signs in both \eqref{J} and \eqref{bkg}, which is due to our non-standard convention.
In equilibrium, the distribution function is given by Fermi-Dirac distribution $f_\pm(|p_0|)=\frac{1}{e^{(|p_0|\mp\m)/T}+1}$ with the upper/lower sign for positively/negatively charged LLL states. Higher LL states are massive from the $1+1D$ point of view with mass $\sim\sqrt{nB}$, thus their contribution are exponentially suppressed $\sim e^{-\sqrt{nB}/T}$.

Since \eqref{ckt_LL} is valid to $O(\pd_X)$, we seek solution of $j^\m$ order by order in gradient:
\begin{align}\label{j_grad}
  \d j^\m=\d j^\m_{(0)}+\d j^\m_{(1)}+\cdots,
\end{align}
with the subscript indicating order of gradient. We use $\d$ to distinguish the induced $j^\m$ from the background one. Terms of $O(\pd_X^2)$ are beyond the accuracy of \eqref{ckt_LL}. Substituting \eqref{j_grad} into \eqref{ckt_LL}, we obtain to order $O(\pd_X^0)$ and $O(\pd_X)$ respectively
\begin{align}\label{EOM0}
  &D_i\d j^i_{(0)}=0,\no
  &D_i\d j^0_{(0)}+2\e^{ijk}p_j\d j^k_{(0)}=0,\no
  &p_0\d j^0_{(0)}+p_i\d j^i_{(0)}=0,\no
  &-p_0\d j^i_{(0)}-p_i\d j^0_{(0)}+\frac{1}{2}\e^{ijk}D_j\d j^k_{(0)}=0,
\end{align}
and
\begin{align}\label{EOM1}
  &D_i\d j^i_{(1)}=S,\no
  &D_i\d j^0_{(1)}+2\e^{ijk}p_j\d j^k_{(1)}=V^1_{i},\no
  &p_0\d j^0_{(1)}+p_i\d j^i_{(1)}=0,\no
  &-p_0\d j^i_{(1)}-p_i\d j^0_{(1)}+\frac{1}{2}\e^{ijk}D_j\d j^k_{(1)}=V^2_{i},
\end{align}
with the right hand side defined as
\begin{align}\label{SV}
  &S=-\(\d\D_0j^0+\d\D_ij^i+\pd_0\d j^0_{(0)}+\pd_i\d j^i_{(0)}\),\no
  &V^1_{i}=-\(\d\D_0j^i+\d\D_ij^0+\pd_0\d j^i_{(0)}+\pd_i\d j^0_{(0)}\),\no
  &V^2_{i}=-\(\frac{1}{2}\e^{ijk}\d\D_jj^k+\frac{1}{2}\e^{ijk}\pd_j\d j^k_{(0)}\).
\end{align}
We have defined $D_i=-\frac{\pd}{\pd p_j}F_{ij}$ and $\d\D_\m=-\frac{\pd}{\pd p_\n}f_{\m\n}$. We also use the short-hand notation $\pd_\m=\pd_\m^X$.
The structure of the equations are quite informative: \eqref{EOM0} and \eqref{EOM1} can be viewed as equations for $\d j^\m_{(0)}$ and $\d j^\m_{(1)}$ respectively. The only difference is that the former are homogeneous and the latter are inhomogeneous. The homogeneous equations can't uniquely determine $\d j^\m_{(0)}$. The source of inhomogeneous equations involves perturbations and the undetermined $\d j^\m_{(0)}$. Nevertheless, \eqref{EOM0} and \eqref{EOM1} can still be solved thanks to the over-determinancy of the equations, which we will elaborate in the next section.

Before closing this section, we verify that the first equation of \eqref{EOM1} is consistent with the anomalous Ward identity. We integrate the equation over four momentum and reorganize it as.
\begin{align}
  \int d^4p \(\pd_\m\d j^\m_{(0)}\)=-\int d^4p\(D_i\d j^i_{(1)}+\d\D_0j^0+\d\D_3j^3\).
\end{align}
By our non-standard convention, we identify the left hand side (LHS) as $\pd_\m J^\m$. For the right hand side (RHS), the first term becomes boundary terms upon integration over transverse momentum
\begin{align}
  \int d^2p_TD_i\d j^i_{(0)}=\int d^2p_T\frac{\pd}{\pd p_M}B\e^{MN}\d j^N_{(0)}=0,
\end{align}
with the capital Roman letters run over indices in the plane perpendicular to the background magnetic field $M,N=1,2$. The second terms can be written explicitly as
\begin{align}
  \int d^4p\[\(\frac{\pd}{\pd p_M}f_{0M}+\frac{\pd}{\pd p_3}f_{03}\)j^0+\(\frac{\pd}{\pd p_0}f_{30}+\frac{\pd}{\pd p_M}f_{3M}\)j^3\].
\end{align}
The terms involving $\frac{\pd}{\pd p_M}$ vanish for the same reason as above. Using \eqref{bkg} and including contribution from both positively and negatively charged LLL states, we can combine the remaining terms as
\begin{align}\label{sat_aWI}
  &\int d^4p\(\frac{\pd}{\pd p_3}-\frac{\pd}{\pd p_0}\)f_{03}\frac{2}{(2\pi)^3}exp(-p_T^2/B)\d(p_0+p_3)\(f_+(|p_0|)-f_-(|p_0|)\)\no
  =&\frac{E_3B}{(2\pi)^2}\(\int_0^\infty dp_0\(-\frac{\pd}{\pd p_0}\)f_+(|p_0|)-\int_{-\infty}^0dp_0\(-\frac{\pd}{\pd p_0}\)f_-(|p_0|)\)\no
  =&\frac{E_3B}{(2\pi)^2}\(f_+(0)+f_-(0)\)=\frac{E_3B}{(2\pi)^2}.
\end{align}
Therefore, we reproduce anomalous Ward identity. It also shows the anomalous Ward identity is saturated by zeroth order solution in the regime of strong magnetic field.

\section{Self-energy from solutions to kinetic equations}

In this section, we present solutions to \eqref{EOM0} and \eqref{EOM1}, which allow us to extract components of self-energy in different kinematics, which in fact correspond to different states. After presenting three simple solutions, we will show no more solution is possible. We will argue this is an artifact of the collisionless limit.

\subsection{Parallel $E$ field: chiral magnetic wave}

We begin with the case of parallel $E$ field, which can be induced by either $a_0(t,x_3)$ or $a_3(t,x_3)$. This case can be simplified by noting that the longitudinal motion of LL states is classical. In the LLL approximation, parallel electric field only induces redistribution of LLL states. Since we know LLL state satisfies homogeneous equation \cite{Lin:2019fqo}, we expect $\d j^\m_{(0)}$ proportional to LLL state and $\d j^\m_{(1)}=0$. It follows that \eqref{EOM0} are satisfied automatically. In order for \eqref{EOM1} to hold, we need to require the inhomogeneous terms to vanish: $S=V^1_{i}=V^2_{i}=0$. It gives the following constraint on $\d j^\m_{(0)}$
\begin{align}\label{SV0}
  f_{03}\(\frac{\pd}{\pd p_3}-\frac{\pd}{\pd p_0}\)j^0-\(\pd_0+\pd_3\)\d j^0_{(0)}=0.
\end{align}
In arriving at \eqref{SV0}, we have used property of LLL state: $\d j^0_{(0)}=\d j^3_{(0)}$ and assumed all $\pd_M$ to vanish because the field perturbation $f_{03}=\pd_0a_3-\pd_3a_0$ is independent on $x_T$. \eqref{SV0} can be solved easily in momentum space as
\begin{align}\label{sol_epara}
  &\d j^0_{(0)}=\d j^3_{(0)}=\frac{2}{(2\pi)^3}exp(-p_T^2/B)\d(p_0+p_3)f'(p_0)\frac{q_3}{q_0+i\e+q_3}a_0, \no
  &\d j^0_{(0)}=\d j^3_{(0)}=-\frac{2}{(2\pi)^3}exp(-p_T^2/B)\d(p_0+p_3)f'(p_0)\frac{q_0}{q_0+i\e+q_3}a_3,
\end{align}
for perturbations $a_0(t,x_3)$ and $a_3(t,x_3)$ respectively. Again $f(p_0)$ can be $f_\pm(|p_0|)$ for solutions corresponding to positively and negatively charged LLL states. We have made the substitution $q_0\to q_0+i\e$ so that the solution corresponds to retarded response.

Integrating the solution \eqref{sol_epara} over four momentum and using the following identity
\begin{align}
  \int_0^\infty dp_0f'_+(|p_0|)-\int_{-\infty}^0dp_0f'_-(|p_0|)=-f_+(0)-f_-(0)=-1,
\end{align}
we obtain the following retarded self-energy components from \eqref{Pi_p}
\begin{align}\label{Pi_epara}
  &\Pi^{00}(q_0,q_3)=\Pi^{30}(q_0,q_3)=-\frac{e^3B}{(2\pi)^2}\frac{q_3}{q_0+i\e+q_3},\no
  &\Pi^{03}(q_0,q_3)=\Pi^{33}(q_0,q_3)=\frac{e^3B}{(2\pi)^2}\frac{q_0}{q_0+i\e+q_3}.
\end{align}
We have reinstated powers of $e$ in the above.
Note that \eqref{Pi_epara} is independent of temperature and chemical potential. This is because the integration over $p_0$ only picks up boundary terms at $p_0=0$, similar to \eqref{sat_aWI}.
We can verify \eqref{Pi_epara} indeed satisfies \eqref{aWI}. The pole of $\Pi^{\m\n}$ gives dispersion relation $q_0+q_3=0$ of collective excitation of the chiral medium. It corresponds to a wave propagating with speed of light in the $x_3$ direction. In fact, this is nothing but chiral magnetic wave in the limit of strong magnetic field \cite{Kharzeev:2010gd}.

It is also interesting to note $\Pi^{\m\n}$ is not symmetric with respect to Lorentz indices. The reason is that the state consisting of right-handed fermions are not parity invariant. Applying parity transformation on \eqref{Pi_epara}, we obtain components of self-energy for medium consisting of left-handed fermions
\begin{align}\label{Pi_parity}
  &\Pi^{00}(q_0,q_3)=-\Pi^{30}(q_0,q_3)=\frac{e^3B}{(2\pi)^2}\frac{q_3}{q_0+i\e-q_3},\no
  &-\Pi^{03}(q_0,q_3)=\Pi^{33}(q_0,q_3)=\frac{e^3B}{(2\pi)^2}\frac{q_0}{q_0+i\e-q_3}.
\end{align}
These results can also be obtained from explicit solution of chiral kinetic theory for left-handed medium, which we collect in Appendix A. Adding up contributions from both left-handed and right handed fermions, we obtain the following components of self-energy for parity invariant state
\begin{align}\label{Pi_inv}
  &\Pi^{00}(q_0,q_3)=\frac{e^3B}{(2\pi)^2}\frac{2q_3^2}{(q_0+i\e)^2-q_3^2},\no
  &\Pi^{03}(q_0,q_3)=\Pi^{30}(q_0,q_3)=-\frac{e^3B}{(2\pi)^2}\frac{2q_0q_3}{(q_0+i\e)^2-q_3^2},\no
  &\Pi^{33}(q_0,q_3)=\frac{e^3B}{(2\pi)^2}\frac{2q_0^2}{(q_0+i\e)^2-q_3^2}.
\end{align}
These agree with field theoretic results in the LLL approximation up to an overall factor of $e^{-q_T^2/B}$ \cite{Fukushima:2011nu}. The exponential factor at least is $O(\pd_X^2)$, which lies beyond the accuracy of our current chiral kinetic equation.

\subsection{Static perpendicular $E$ field: drift state}

Next we consider the case of static perpendicular $E$ induced by $a_0(x_T)$. We begin by solving for $\d j^0_{(1)}$ and $\d j^3_{(1)}$ from the third and $i=3$ component of the fourth equations of \eqref{EOM1} to obtain
\begin{align}\label{j03_1}
  \begin{pmatrix}
  \d j^0_{(1)}\\
  \d j^3_{(1)}
  \end{pmatrix}=\frac{1}{p_0^2-p_3^2}
  \begin{pmatrix}
    -p_0p_M\d j^M_{(1)}-p_3\frac{1}{2}\e^{MN}D_M\d j^N_{(1)}\\
    p_3p_M\d j^M_{(1)}+p_0\frac{1}{2}\e^{MN}D_M\d j^N_{(1)}
  \end{pmatrix}.
\end{align}
%Plugging the above into $i\ne3$ components of the fourth equations of \eqref{EOM1}, we have
%\begin{align}\label{long}
%  &-p_0\d j^M_{(1)}-p_M\frac{-p_0p_M\d j^M_{(1)}-p_3\frac{1}{2}\e^{MN}D_M\d j^N_{(1)}}{p_0^2-p_3^2}+\frac{1}{2}\e^{MN}D_N\frac{p_3p_M\d j^M_{(1)}+p_0\frac{1}{2}\e^{MN}D_M\d j^N_{(1)}}{p_0^2-p_3^2}\nonumber
%  &=\frac{1}{2}\e^{MN}\pd_Na_0\frac{\pd}{\pd p_0}j^0-\frac{1}{2}\e^{MN}\pd_N\d j^3_{(0)}.
%\end{align}
We proceed with the following ansatz
\begin{align}\label{ansatz}
  \d j^M_{(1)}\propto exp(-p_T^2/B)\d(p_0+p_3),
\end{align}
which converts \eqref{EOM1} to the following equivalent equations
\begin{align}\label{equiv}
  &\frac{B\d j^M_{(1)}}{2(p_0+p_3)}-p_0\d j^M_{(1)}=-\frac{1}{2}\e^{MN}E_N\frac{\pd}{\pd p_0}j^0-\frac{1}{2}\e^{MN}\pd_N\d j^3_{(0)},\no
  &2\e^{MN}p_M\d j^N_{(1)}=-\frac{2E_Mp_M}{B}j^0,\no
  &\frac{\e^{MN}\d j^N_{(1)}B}{p_0+p_3}+2\e^{MN}\d j^N_{(1)}p_3=E_M\frac{\pd}{\pd p_0}j^0+\pd_M\d j^0_{(0)},
\end{align}
with $E_N=-\pd_Na_0$ being the perpendicular electric field perturbation.
\eqref{equiv} adopts the following solution
\begin{align}\label{sol_eperp}
  &\d j^0_{(0)}=\d j^3_{(0)}=a_0\(-\d'(p_0+p_3)-\frac{2p_0}{B}\d(p_0+p_3)\)exp(-p_T^2/B)f(p_0),\no
  &\d j^M_{(1)}=\frac{\e^{MN}E_N}{B}\d(p_0+p_3)exp(-p_T^2/B)f(p_0),\no
  &\d j^0_{(1)}=\d j^3_{(1)}=\frac{\e^{MN}p_ME_N}{B}\d'(p_0+p_3)exp(-p_T^2/B)f(p_0).
\end{align}
Integrating the solution over four momentum and using the following integrals
\begin{align}
  &\int_0^\infty dp_0p_0f_+(|p_0|)-\int_{-\infty}^0 dp_0p_0f_-(|p_0|)=\frac{\pi^2T^2}{6}+\frac{\m^2}{2},\no
  &\int_0^\infty dp_0f_+(|p_0|)-\int_{-\infty}^0 dp_0f_-(|p_0|)=\m,\nonumber
\end{align}
we find the following components of self-energy after reinstating powers of $e$
\begin{align}\label{Pi_eperp}
  &\Pi^{00}(q_M)=\Pi^{30}(q_M)=-\frac{e^2}{(2\pi)^2}\(eB+\frac{\pi^2T^2}{3}+\m^2\),\no
  &\Pi^{M0}(q_M)=-\frac{e^2}{(2\pi)^2}i\e^{MN}q_N\m.
\end{align}
Note that $\d j^0_{(1)}$ and $\d j^3_{(1)}$ are odd function of $p_T$, which vanishes upon integration over $p_T$, thus do not contribute to self-energy.
Note that unlike the case of parallel electric field, the case of perpendicular electric field gives rise to medium dependent self-energy components. In particular the medium dependent terms in $\Pi^{00}$ and $\Pi^{30}$ would not appear in the static limit $q_0\to0$ of \eqref{Pi_epara}. The difference can be understood as follows: the solutions from parallel and perpendicular electric fields correspond to different state: parallel electric field leads to redistribution of LLL states, while the perpendicular electric field leads to a drift state, with the medium drifting with a velocity orthogonal to both $E_M$ and $B$. In the drift state, Hall current is expected and is consistent with $\Pi^{M0}$ above. $\Pi^{00}$ and $\Pi^{30}$ give deviation of charge density and current density of the drift state from those of the background. Interestingly the deviation coincides with $00$ component of the self-energy in chiral medium without background magnetic field \cite{Akamatsu:2013pjd}.

%In fact, the disagreement of \eqref{Pi_epara} and  \eqref{Pi_eperp} is an artifact of the collisionless limit we work in. To develop the medium velocity in the drift state, interaction between LL states is needed. In the absence of interaction, the solution corresponding to drift state can be found, but cannot be dynamically realized from unperturbed equilibrium state. %Therefore \eqref{Pi_eperp} should be understood as charge and current in drift state, rather than response to perpendicular electric field.

We can gain further insight of the drift state by ``boosting'' the equilibrium state. It is not difficult to see that boosting the equilibrium medium to a velocity $-\frac{\e^{MN}E_N}{B}$, we have then orthogonal electric field $E_M$ and background magnetic field $B$ \footnote{The magnetic field in boosted frame is $\sqrt{B^2-E^2}$, whose deviation from background is negligible to linear order in $E_M$.}. To describe it more quantitatively, we use the covariant form of the background solution
\begin{align}\label{cov_bkg}
  j^\m=\frac{2}{(2\pi)^3}\d(p\cdot(u+b))e^{-p_T^2/B}f(p\cdot u)\(u+b\)^\m,
\end{align}
which generalizes the solution in medium frame to arbitrary frame. We verify in appendix B that it indeed satisfies covariant chiral kinetic equations to the lowest order in gradient. Here $u^\m$ and $b^\m$ denote fluid velocity and magnetic direction and $p_T^2=-p^2+(p\cdot u)^2-(p\cdot b)^2$. $u^\m$ and $b^\m$ are orthogonal to each other $u\cdot b=0$. In medium frame, we have $u^\m=(1,0,0,0)$ and $b^\m=(0,0,0,1)$. Under the boost, we have $\d u^M=\frac{\e^{MN}E_N}{B}$ and $\d b=0$. It is easy to see that $\d u^M$ leads to $\d j^M_{(1)}$. The remaining corrections are not from covariance and are only present in $\d j^0$ and $\d j^3$. At zeroth order $\d j^0_{(0)}$ and $\d j^3_{(0)}$ come from the fact that the electric field perturbation is not constant but $x_T$ dependent. In fact, $a_0$ in \eqref{sol_eperp} should be interpreted as $a_0\sim E_M/\pd_M$, thus the zeroth order correction characterizes redistribution of LL states in response to perturbation. Similar correction to zeroth order solution is also present in chiral kinetic theory without background field \cite{Son:2012zy}. At the first order $\d j^0_{(1)}$ and $\d j^3_{(1)}$ are entirely determined by $\d j^M_{(1)}$ from \eqref{j03_1}, which as we discussed above is not sensitive to $x_T$ dependence of the perturbation. Note that $\d'(p_0+p_3)f(p_0)=-\d(p_0+p_3)f'(p_0)$. It is suggestive to interpret $\d j^0_{(1)}$ and $\d j^3_{(1)}$ as modification of distribution function $f(p_0)\to f(p_0)-\frac{\e^{MN}p_ME_N}{B}f'(p_0)=f(p_0-\frac{\e^{MN}p_ME_N}{B})$, or $p_0\to p_0-\frac{\e^{MN}p_ME_N}{B}$. This is analogous to side-jump effect in momentum in the absence of background field \cite{Chen:2014cla,Hidaka:2016yjf}. Note that since our background solution is homogeneous in coordinate, possible jump in coordinate is not visible from our comparison. We should not confuse the frame vector frequently used in description of side-jump with the fluid velocity $u^\m$. The latter is needed to define magnetic field in the background.
%It is also interesting to note that $\d j^0_{(0)}$ and $\d j^3_{(0)}$ could be generated if we take $p_p\to p_0+a_0$ in $e^{-p_T^2/B}f(p\cdot u)$. Taking variation with respect to $a_0$ we reproduce $\d j^0_{(0)}$ and $\d j^3_{(0)}$.

\subsection{Static perpendicular $B$ field: tilted state}

We turn to the case of static perpendicular magnetic field induced by $a_3(x_T)$. The analysis is similar to the previous subsection. We will not spell out details.  With the ansatz $\d j^M_{(1)}\propto exp(-p_T^2/B)\d(p_0+p_3)$, we obtain the following solution
\begin{align}\label{sol_bperp}
  &\d j^0_{(0)}=\d j^3_{(0)}=\(-\frac{2p_0}{B}\d(p_0+p_3)exp(-p_T^2/B)f(p_0)\)a_3,\no
  &\d j^0_{(1)}=\d j^3_{(1)}=\frac{p_MB^\perp_M}{B}\d'(p_0+p_3)exp(-p_T^2/B)f(p_0),\no
  &\d j^M_{(1)}=\frac{B^\perp_M}{B}\d(p_0+p_3)exp(-p_T^2/B)f(p_0),
\end{align}
with $B_M^\perp=-\e^{MN}\pd_Na_3$ being the perpendicular magnetic field. It gives rise to the following components of self-energy
\begin{align}\label{Pi_bperp}
  &\Pi^{03}(q_M)=\Pi^{33}(q_M)=-\frac{e^2}{(2\pi)^2}\(\frac{\pi^2T^2}{3}+\m^2\),\no
  &\Pi^{M3}(q_M)=-\frac{e^2}{(2\pi)^2}i\e^{MN}q_N\m.
\end{align}
Comparing \eqref{Pi_bperp} with the static limit of \eqref{Pi_epara}, which vanishes identically, we see the difference is also medium dependent. %Similar to the case of perpendicular electric field leading to solution of drift state, \eqref{sol_bperp} should not be understood in the sense of retarded response, because the magnetic field perturbation is static.
The presence of $B_\perp$ can also be understood as tilt of the background. We can see $\Pi^{M3}$ gives precisely chiral magnetic effect for right-handed fermions due to $B_\perp$:
\begin{align}
  J^M=\Pi^{M3}a_3=-\frac{e^2}{(2\pi)^2}\e^{MN}\pd_Na_3\m=\frac{e^2}{(2\pi)^2}\m B_M.
\end{align}
$\Pi^{M3}$ also agrees with components of parity odd self-energy in the absence of background field \cite{Son:2012zy}, which is responsible for chiral magnetic effect.
%The remaining components $\Pi^{03}$ and $\Pi^{33}$ give deviation of charge density and current density of the tilted state from those of the background.

It is interesting to note that this particular components of self-energy actually gives rise to chiral plasma instability \cite{Akamatsu:2013pjd}. We can see some trace from the backreaction of the induced chiral magnetic current to the electromagnetic field. The induced magnetic field $\D B$ can be found by solving the Maxwell equation
\begin{align}
  \(\nabla\times \D B\)^M=J^M,
\end{align}
with the solution given by $\D B_i=\frac{e^2}{(2\pi)^2}\m a_3\d_{i3}$. It enhances the original perturbation of perpendicular magnetic field. The mechanism of enhancement seems independent of the background magnetic field. However, to answer the question dynamically, we need to know the self-energy away from the static limit.

Let us again compare the tilted state with the equilibrium state rotated in such a way that the background magnetic field coincides with that of the tilted state \footnote{To linear order in the perturbation, the magnitude of the magnetic field $\sqrt{B^2+B_\perp^2}=B$ does not change. Only the direction changes.}. We can use the covariant form of background solution \eqref{cov_bkg} with $\d b^M=B^\perp_M/B$ and $\d u=0$. $\d b^M$ gives precisely $\d j^M_{(1)}$ from the covariant factor $(u+b)^\m$. The remaining difference between tilted state and equilibrium state is in $\d j^0$ and $\d j^3$. The tilted state is not entirely equivalent to rotated equilibrium state because of the $x_T$ dependence of perpendicular magnetic field. The $x_T$ dependence leads to the the difference in $\d j^0_{(0)}$ and $\d j^3_{(0)}$. On the contrary, $\d j^0_{(1)}$ and $\d j^3_{(1)}$ are fixed by $\d j^M_{(1)}$ through \eqref{j03_1}, thus are not sensitive to $x_T$ dependence of the perturbation. Using $\d'(p_0+p_3)f(p_0)=-\d(p_0+p_3)f'(p_0)$, we interpret $\d j^0_{(1)}$ and $\d j^3_{(1)}$ as modification of distribution function: $f(p_0)\to f(p_0)-\frac{p_MB_M}{B}f'(p_0)=f(p_0-\frac{p_MB_M}{B})$, or $p_0\to p_0-\frac{p_MB_M}{B}$. Like in the case of drift state, it is suggestive to interpret the modification of distribution function as analog to side-jump effect in momentum in the absence of background field \cite{Chen:2014cla,Hidaka:2016yjf}. Since our background solution is homogeneous in coordinate, possible jump in coordinate is not visible.
We stress that this is a new effect due to the background field: conventional side-jump is manifested through boost, in our case side-jump can be manifested through both boost $\d u$ and rotation $\d b$, as we see in both drift state and tilted state respectively.

\subsection{No more solutions}

Finally, we look for solution for more general perturbations without using the ansatz. Note that $S=V^1_{3}$ and $\frac{1}{2}\e^{MN}V^1_{N}=V^2_{M}$. We can eliminate redundant equations in \eqref{EOM1} to obtain
\begin{align}\label{EOM2}
  &D_M\d j^M_{(1)}=2\e^{MN}p_M\d j^N_{(1)},\no
  &\(\frac{1}{2}\e^{MN}D_N+p_M\)\(\d j^0_{(1)}-\d j^3_{(1)}\)+(p_0+p_3)\d j^M_{(1)}=0,\no
  &2\e^{MN}p_M\d j^N_{(1)}=-\(\(\d\D_0+\d\D_3\)j^0+\(\pd_0+\pd_3\)\d j^0_{(0)}\),\no
  &(p_0-p_3)\e^{MN}\d j^N_{(1)}+\(\frac{1}{2}D_M+\e^{MN}p_N\)\(\d j^0_{(1)}+\d j^3_{(1)}\)=-\(\d\D_Mj^0+\pd_M\d j^0_{(0)}\).
\end{align}
%Here the first two equations are homogeneous and the last two equations are inhomogeneous.
To proceed, we define $A_M$ by pulling out a factor of $e^{-p_T^2/B}$ from $\d j^M_{(1)}$: 
\begin{align}\label{decomp}
  \d j^M_{(1)}=e^{-p_T^2/B}A_M.%\(q_MA_1+\e^{MN}q_NA_2\).
\end{align}
%Note that this basis is complete in 2D. We have pulled out a factor of $e^{-p_T^2/B}$ for convenience.
$A^M$ are functions of $p$ and $q$. The solutions for static perpendicular electric and magnetic fields correspond to $A_M$ being independent of $p_M$. Plugging \eqref{decomp} into \eqref{EOM2} and divide out common factor $e^{-p_T^2/B}$, we obtain
\begin{subequations}
  \begin{align}
  &\e^{MN}\frac{\pd}{\pd p_N}A_M=0,\label{eq1}\\
  &-\frac{1}{p_0-p_3}\(2p_M-\frac{B}{2}\frac{\pd}{\pd p_M}\)\frac{B}{2}\frac{\pd}{\pd p_K}A_K+(p_0+p_3)A_M=0,\label{eq2}\\
    &2\e^{MN}p_MA_N=f_{30}\d(p_0+p_3)f'(p_0)-\frac{2p_K}{B}(f_{0K}+f_{3K})\d(p_0+p_3)-(\pd_0+\pd_3)\d(p_0+p_3)g(p_0),\label{eq5}\\
    &(p_0-p_3)\e^{MN}A_N+\frac{-1}{p_0+p_3}\frac{B}{2}\e^{MN}\frac{\pd}{\pd p_N}\[\(2p_K-\frac{B}{2}\frac{\pd}{\pd p_K}\)A_K\]\no
    &=\(f_{M0}\frac{\pd}{\pd p_0}+f_{M3}\frac{\pd}{\pd p_3}\)\(\d(p_0+p_3)f(p_0)\)+\pd_M\d(p_0+p_3)g(p_0).\label{eq6}
  \end{align}
\end{subequations}
We have defined $\d j^0_{(0)}=\d(p_0+p_3)e^{-p_T^2/B}g(p_0)$. We already know $\d j^\m_{(0)}$ corresponds to redistribution of LLL states, so $g$ has to be a function of $p_0$ only. Below we will show this is not possible except for the special cases presented in the above subsections.
We first apply $\e^{MN}\frac{\pd}{\pd p_N}$ to \eqref{eq2} and use \eqref{eq1} to arrive at
\begin{align}\label{eq3}
  p_M\e^{MN}\frac{\pd}{\pd p_N}\frac{\pd}{\pd p_K}A_K=0.
\end{align}
We can also apply $\e^{MN}p_N$ to \eqref{eq2} and use \eqref{eq3} to obtain
\begin{align}\label{eq4}
  (p_0+p_3)\e^{MN}p_MA_N=0.
\end{align}
We then multiply \eqref{eq6} by $p_M$ to obtain:
\begin{align}
  &(p_0-p_3)p_M\e^{MN}A_N+\frac{-1}{p_0+p_3}\frac{B}{2}\e^{MN}p_M\frac{\pd}{\pd p_N}\(2p_KA_K-\frac{B}{2}\frac{\pd}{\pd p_K}A_K\)\no
  =&p_M\(f_{M0}\frac{\pd}{\pd p_0}+f_{M3}\frac{\pd}{\pd p_3}\)\(\d(p_0+p_3)f(p_0)\)-p_M\pd_M\d(p_0+p_3)g(p_0).
\end{align}
Using \eqref{eq4} to simplify the first term and using \eqref{eq3} to eliminate the second term in the round bracket on the LHS, we arrive at
\begin{align}\label{eq7}
  &2p_0\e^{MN}p_MA_N+\frac{-B}{p_0+p_3}\[\e^{MN}p_MA_N+\e^{MN}p_Mp_K\frac{\pd}{\pd p_N}A_K\]=\no
  &p_M\(f_{M0}\frac{\pd}{\pd p_0}+f_{M3}\frac{\pd}{\pd p_3}\)\(\d(p_0+p_3)f(p_0)\)-p_M\pd_M\d(p_0+p_3)g(p_0).
\end{align}
The second term in the square bracket can be further simplified using the following identity
\begin{align}\label{eid}
  \e^{MN}p_K+\e^{NK}p_M+\e^{KM}p_N=0.
\end{align}
It follows that
\begin{align}
  \e^{MN}p_Mp_K\frac{\pd}{\pd p_N}A_K=-\(\e^{NK}p_M+\e^{KM}p_N\)p_M\frac{\pd}{\pd p_N}A_K=\e^{MN}p_Mp_K\frac{\pd}{\pd p_K}A_N,
\end{align}
where we have used \eqref{eq1} and relabeled indices in the second equality. We can then rewrite the square bracket of \eqref{eq7} as
\begin{align}
  \[\e^{MN}p_MA_N+\e^{MN}p_Mp_K\frac{\pd}{\pd p_N}A_K\]=\e^{MN}p_M\(1+\frac{\pd}{\pd p_K}\)A_N=\e^{MN}\frac{\pd}{\pd p_K}p_MA_N.
\end{align}
With this, we arrive at the following simple form of \eqref{eq7}
\begin{align}\label{eq8}
  &\(2p_0+\frac{-B}{p_0+p_3}\)\e^{MN}p_MA_N=\no
  &p_M\(f_{M0}\frac{\pd}{\pd p_0}+f_{M3}\frac{\pd}{\pd p_3}\)\(\d(p_0+p_3)f(p_0)\)-p_M\pd_M\d(p_0+p_3)g(p_0).
\end{align}
We can now plug \eqref{eq5} into the above and compare coefficients of $\d(p_0+p_3)$ and $\d'(p_0+p_3)$ to determine $g(p_0)$. Note that the coefficients have to be matched separately rather than using $\d'(p_0+p_3)=-\frac{\d(p_0+p_3)}{p_0+p_3}$, which involves dropping of boundary terms and is not always justified. We end up with two expressions for $g$:
\begin{align}\label{f0}
  &g=\frac{\tilde{f}_{30}f'(p_0)}{i(q_0+q_3)},\no
  &g=\frac{(2p_0\tilde{f}_{30}-2p_M\tilde{f}_{M0})f'(p_0)-\frac{2p_0}{B}\(2p_M\tilde{f}_{0M}+2p_M\tilde{f}_{3M}\)f(p_0)}{2i\(p_0(q_0+q_3)-p_Mq_M\)},
\end{align}
with $\tilde{f}_{\m\n}=i(q_\m a_\n-q_\n a_\m)$.
It is easy to verify that \eqref{f0} include all three cases we discussed above: in the case of parallel electric field, two expressions of \eqref{f0} give the same result; in the case of static perpendicular electric or magnetic field, only the second expression should be used. In all three cases, $g$ is independent of $p_T$. This is a necessary condition for $\d j^0_{(0)}$ to be a valid zeroth order solution as stressed above. However, any other field perturbations would not allow for a $p_T$ independent $g$ thus no more solution can be found.

The lack of nontrivial solution may sound odd. Indeed, it is actually an artifact of the collisionless limit we work in. In the absence of interaction between LL states, the dynamics of LL states is restricted to classical longitudinal motion. This can be induced by parallel electric field leading to chiral magnetic wave. Perpendicular electric or magnetic field necessarily leads to quantum transition between LL states. However, this cannot occur without interaction. The only possible solution is static ones in which no dynamics is involved. In other words, although these static solutions can be found, solutions corresponding dynamical realization of these states is not possible in the absence of collision. The collisionless limit also lies behind the disagreement of the static limit of \eqref{Pi_epara} and \eqref{Pi_eperp}. We expect a consistent limit will be reached by including collision. We leave it for future work.

\section{Summary and Outlook}

By using chiral kinetic theory with Landau level states, we studied photon self-energy in magnetized chiral plasma from response to electromagnetic field perturbations. In the regime of strong magnetic field, we studied the response of chiral plasma to three different field perturbations: parallel electric field, static perpendicular electric and static perpendicular magnetic fields. They give rise to components of self-energy in different kinematics. The three perturbations lead to chiral magnetic wave, drift state and tilted state respectively. From the case of chiral magnetic wave, we obtain self-energy components, which are in agreement with field theoretic results up to the accuracy of the chiral kinetic theory. From the cases of drift state and tilted state, we obtain components of self-energy in the static case. We also compared the solutions of drift state and tilted state with boosted and rotated background solution respectively. The difference is understood from the spatial dependence of of the perturbations.

We further showed no solution can be found in response to other more general perturbations. We argued it is an artifact of the collisionless limit we work. Without collisions, quantum transition between LL states is not possible but only classical motion of LL state is allowed. As a result, drift state and tilted state cannot be realized dynamically but can only be found as static solutions. To study more general perturbations, it is crucial to introduce collision. It can be done by promoting photon as a dynamical field, which mediates interaction between LL states. It would also allow us to study self-energy of LL states and photon in a systematic way. We hope to address these in the future.

\begin{acknowledgments}
S.L. is grateful to Koichi Hattori, Defu Hou, Igor Shovkovy and Di-Lun Yang for useful discussions. He also thanks Yukawa Institute of Theoretical Physics for hospitality and the workshop ``Quantum kinetic theories in magnetic and vortical fields'' for providing a stimulating environment during the final stage of this work. This work is in part supported by NSFC under Grant Nos 11675274 and 11735007.
\end{acknowledgments}

\appendix

\section{Chiral kinetic equations for left handed fermions}

The chiral kinetic equation for left handed fermions can be derived from the equation of motion for the corresponding Wigner function $W$:
\begin{align}\label{Wigner_eom}
  &\(\frac{1}{2}\D_\m-ip_\m\)\bar{\s}^\m W=0,
  &\(\frac{1}{2}\D_\m+ip_\m\)W\bar{\s}^\m =0.
\end{align}
The difference with counterpart of right handed fermions is $\s^\m\to\bar{\s}^\m$. We then decompose the Wigner function into components $j^\m$ as
\begin{align}
  W=j^01+j^i\s_i.
\end{align}
This decomposition keeps the integral representation of current \eqref{J} the same for left handed fermions. The chiral kinetic equations for components follow immediately from \eqref{Wigner_eom}
\begin{align}
  &\D_0j^0-\D_ij^i=0,\no
  &p_0j^0-p_ij^i=0,\no
  &\D_0j^i-\D_ij^0-2\e^{ijk}p_jj^k,\no
  &-p_0j^i+p_ij^0-\frac{1}{2}\e^{ijk}\D_jj^k=0.
\end{align}
They are obtainable from the counterpart of right handed fermions by the replacement $\D_i\to-\D_i$ and $p_i\to -p_i$. It follows that the contributions of left handed fermions and right handed fermions to self-energy are related by the replacement $q_i\to-q_i$, which agrees with what we used in the text.

\section{Covariance of the background solution}

In this appendix, we show the covariance of the background solution \eqref{cov_bkg}. We first write down the covariant chiral kinetic equations
\begin{subequations}
\begin{align}
  &\D_\m j^\m=0,\label{e1}\\
  &-\D_\m j_\n+\D_n j_\m+2\e_{\m\n\r\s}p^\r j^\s=0,\label{e2}\\
  &p_\m j^\m=0,\label{e3}\\
  &p_\m j_\n-p_\n j_\m+\frac{1}{2}\e_{\m\n\r\s}\D^\r j^\s=0,\label{e4}
\end{align}
\end{subequations}
with $\D_\m=\pd_\m-\frac{\pd}{\pd p_\n}\(F_{\m\n}+f_{\m\n}\)$. To the lowest order in gradient, we have $\D_\m=-\frac{\pd}{\pd p_\n}F_{\m\n}=\frac{\pd}{\pd p_\n}B\e_{\m\n\r\s}b^\r u^\s$. Here $u^\m$ and $b^\m$ are unit vector corresponding to fluid velocity and magnetic field direction with $u\cdot b=0$. $B\equiv \sqrt{F_{\m\n}F^{\m\n}}$. The solution \eqref{bkg} corresponds to $u^\m=(1,0,0,0)$ and $b^\m=(0,0,0,1)$. In fact, \eqref{e4} is equivalent to \eqref{e2}. This can be shown by multiplying \eqref{e4} by $\e^{\a\b\m\n}$ and using the identity
\begin{align}\label{e_contract}
  \e^{\a\b\m\n}\e_{\m\n\r\s}=-2\(\d_\r^\a\d_\s^\b-\d_\s^\a\d_\r^\b\).
\end{align}
Below we show at the lowest order in gradient \eqref{e1}, \eqref{e3} and \eqref{e4} are indeed satisfied by the following covariant solution
\begin{align}\label{cov_sol}
  j^\m\sim (u+b)^\m\d(p\cdot(u+b))e^{-p_T^2/B}f(p\cdot u),
\end{align}
with $p_T^2=-p^2+(p\cdot u)^2-(p\cdot b)^2$. We first see \eqref{e1} and \eqref{e3} are satisfied by anti-symmetry of indices and on-shell condition:
\begin{align}
  &\D_\m j^\m\sim\frac{\pd}{\pd p_\n}B\e_{\m\n\r\s}b^\r u^\s(u+b)^\m=0,\no
  &p_\m j^\m\sim p\cdot(u+b)\d(p\cdot(u+b))=0.
\end{align}
\eqref{e4} requires some work:
\begin{align}\label{e4_exp}
  &p^\m j^\n-p^\n j^\m+\frac{1}{2}\e^{\m\n\r\s}\D_\r j_\s
  \sim\(p^\m(u+b)^\n-p^\n(u+b)^\m\)\d(p\cdot((u+b))e^{-p_T^2/B}f(p\cdot u)\no
  +&\frac{1}{2}\e^{\m\n\r\s}\frac{\pd}{\pd p_\l}B\e_{\r\s\a\b}b^\a u^\b(u+b)_\s\d(p\cdot(u+b))e^{-p_T^2/B}f(p\cdot u).
\end{align}
The second term can be simplified by noting that $\frac{\pd}{\pd p_\l}$ can pull out $p_\l$, $u_\l$ and $b_\l$. The last two cases always vanish when contracting with $\e_{\r\s\a\b}b^\a u^\b$. Keeping only the $p_\l$ contribution and using the following identity
\begin{align}
  \e^{\m\n\r\s}\e_{\r\s\a\b}=-\(\d_\l^\m\d_\a^\n\d_\b^\s+\d_\a^\m\d_\b^\n\d_\l^\s+\d_\b^\m\d_\l^\n\d_\a^\s-\d_\l^\m\d_\b^\n\d_\a^\s-\d_\b^\m\d_\a^\n\d_\l^\s-\d_\a^\m\d_\l^\n\d_\b^\s\),
\end{align}
we obtain from \eqref{e4_exp}
\begin{align}
  \sim p\cdot(u+b)\(b^\m u^\n-b^\n u^\m\)\d(p\cdot(u+b))e^{-p_T^2/B}f(p\cdot u),
\end{align}
which vanishes by the on-shell condition.

%\bibliographystyle{unsrt}
%\bibliography{}

\end{document}